\documentclass[
floatfix,
aps,pra,amsmath,
twocolumn,
groupaddress,
]{revtex4}

\usepackage{epsfig}
\usepackage{bm}
\usepackage{epsf}
\usepackage{array}

\newcommand{\be}{\begin{equation}}
\newcommand{\ee}{\end{equation}}

\newcommand{\bea}{\begin{eqnarray}}
\newcommand{\eea}{\end{eqnarray}}

\newcommand{\bes}{\begin{split}}
\newcommand{\ees}{\end{split}}

\newcommand{\req}[1]{Eq.~(\ref{#1})}
\newcommand{\reqs}[1]{Eqs.~(\ref{#1})}
\newcommand{\rref}[1]{(\ref{#1})}

\newcommand{\vare}{\varepsilon}

\newcommand{\wc}{\omega_{\rm c}}
\newcommand{\Rc}{R_{\rm c}}
\newcommand{\vF}{v_{\rm F}}
\newcommand{\pF}{p_{\rm F}}
\newcommand{\tautr}{\tau_{\rm tr}}
\newcommand{\tauq}{\tau_{\rm q}}
\newcommand{\hate}{\check{\epsilon}}
\newcommand{\nn}{\mbox{\boldmath $i$}}
\newcommand{\nnf}{\mbox{\boldmath $i$}_{\varphi}}

\newcommand{\p}{\mbox{\boldmath $p$}}
\newcommand{\muB}{\mu_{\rm B}}

\begin{document}

\bibliographystyle{prsty}

\title{Giant Magneto-Oscillations of Electric-Field-Induced Spin Polarization
in 2DEG}

\author{Maxim G. Vavilov}
\affiliation{Department of Applied Physics, Yale University, New
Haven, CT 06520}

\begin{abstract}

We consider a disordered two-dimensional
electron gas with spin-orbit
coupling placed in a perpendicular magnetic field and
calculate the magnitude and direction of
the electric--field--induced spin polarization. We find that in strong magnetic
fields the polarization becomes an oscillatory function of the
magnetic field and that the amplitude of these oscillations is
parametrically larger than the polarization at zero magnetic
field. We show that the enhanced amplitude of the
polarization is a consequence of strong electron--hole asymmetry
in a quantizing magnetic field.

\end{abstract}

\date{30 September, 2005}

\maketitle

\section{Introduction}

One of the main objectives of spintronics~\cite{spintron,ZFS} is
to develop devices, which would control electron spins by electric fields.
A potential implementation of these devices is based
on the magneto-electric effect~\cite{LNE,ALG1,ILGP} in
a two-dimensional electron gas (2DEG) with spin-orbit (SO) coupling.
The spin polarization of 2DEG by dc electric
field, one of the manifestations of the magneto-electric effect,
has recently become a focus of theoretical~\cite{ALG1,Bc2DEG,MSH04,CEM02,IBM03,TDS}
and experimental~\cite{silov,KatoMagn04} investigation. Despite
extensive research~\cite{ALG1,ILGP,MSH04,CEM02,IBM03,Bc2DEG,TDS,silov,KatoMagn04},
the electron--hole asymmetry as the cause of the magneto--electric
effect has not been emphasized.

In this paper, we demonstrate that the magneto--electric
effect in 2DEG is the consequence of electron--hole asymmetry.
Following this observation, we explore potential mechanisms for
enhancement of the electron--hole asymmetry. We find that the
quantization of electron orbital motion in a perpendicular magnetic
field is one of these mechanisms. Particularly, in strong magnetic
field $B$ the polarization induced by an in-plane electric field
oscillates as a function of $B$ with the amplitude of oscillations
larger than the smooth component of the polarization by a huge
factor $\nu=E_{\rm F}/\wc\gg 1$, where $E_{\rm F}=\pF^2/2m^*$ is the
Fermi energy, $\wc=eB/m^*c$ is the cyclotron frequency, $\pF$ is the
Fermi momentum, $e$ and $m^*$  are the charge and effective mass of
electrons, $c$ is the speed of light, $\hbar=1$.

The large parameter $E_{\rm F}/\wc$ is indeed related to the enhancement
of electron--hole asymmetry by magnetic field.
At zero magnetic field, electron scattering rate off disordered
potential is nearly independent of energy. In this case the polarization
is generated by electron--hole asymmetry, which is due to the curvature
of the electron spectrum and is characterized by energy $E_{\rm F}$.
The cyclotron motion of electrons in a perpendicular magnetic
field $B$ results in quantum interference corrections to the
scattering rate off disorder~\cite{Ando,VA03}. These corrections,
periodic in energy, violate the electron--hole
asymmetry on much smaller energy scale $\wc\ll E_{\rm F}$.
We note that some
transport coefficients, such
as the thermoelectric power~\cite{2DEGTP} and the
Coulomb drag transconductance~\cite{GMO},
can be similarly enhanced by magnetic fields.

We derive the quantum kinetic equation for a disordered 2DEG with SO
coupling following the formalism developed in Ref.~\cite{VA03} for
2DEG without SO coupling. We solve this equation and calculate
the spin polarization for a system brought out
of equilibrium by a dc in-plane electric field. The polarization can be
represented as a sum of the smooth and oscillating components,
as illustrated in Fig.~1. At weak magnetic fields,
the oscillatory component is exponentially small, and
only the smooth component remains.
However, the amplitude of oscillatory component
increases as magnetic field increases
and becomes significantly larger than the smooth
component.

The observation of the polarization oscillations  induced
by an electric field seems to be feasible. Indeed, recently,
the  non-equilibrium
spin polarization in zero magnetic field was observed in
experimentally~\cite{KatoMagn04,silov}, and  the possibility of
polarization measurements in strong magnetic fields
was demonstrated in Ref.~\cite{Sih}.

\begin{figure}
\epsfxsize=0.40\textwidth
\centerline{\epsfbox{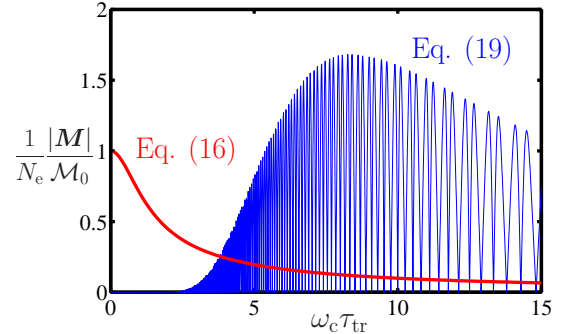}}
\caption{(Color online) The magnitude of the polarization vector
is shown as a function of the magnetic field $B\propto \wc$
at fixed electric field.
The thick smooth line represents the result of \req{noSdH}
for $\lambda_x=\lambda_y$ and $g=0$. The
thin line describes the oscillatory part of the polarization
\req{Mosc} for $\tautr/\tauq=10$, $T=0$, $E_{\rm F}\tautr=1.25\times 10^3$
(period of oscillations is not shown to scale).
}
\label{fig1}
\end{figure}

\section{Qualitative Discussion}

In this section we present a qualitative picture of generation of
spin polarization by in-plane electric field. We show that in
weak magnetic fields, when the electron DoS is energy independent, the
polarization originates due to the dependence of
SO coupling strength on the momentum of electron and hole excitations. On the other
hand, in strong magnetic fields the DoS oscillates as a function
of energy, and therefore the polarization may appear due to the
difference in the number of electron and the number of hole excitations.
We show that the latter mechanism may result in larger
values of the spin polarization.

\begin{figure}
\epsfxsize=0.47\textwidth
\centerline{\epsfbox{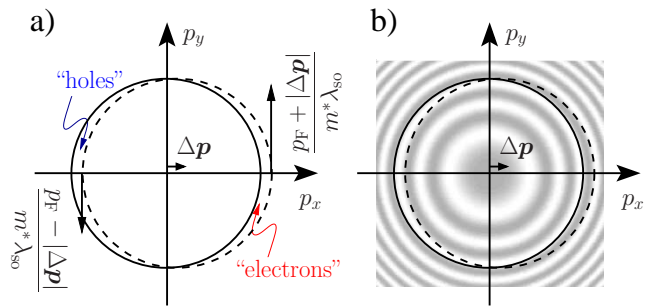}}
\caption{(Color online) a) In equilibrium, electrons occupy all states within Fermi
surface -- a solid circle centered at $\bm{p}=0$. When electric
field is applied, electrons occupy all states within the dashed
circle, centered at $\Delta\bm{p}\propto \bm{j}$. States that become empty are
called hole excitations and states that become occupied are called
electron excitations. The numbers of electron and hole excitations
are equal, and the spin polarization occurs only due to the
difference in the SO coupling of electrons and holes, see
\req{estWF}. The latter is stronger for electrons, that have
larger momentum, than for holes with smaller momentum.
b) In magnetic field, the DoS is modulated, as
shown here by a contour plot. The numbers of
electron and hole excitations are different and the spin
polarization can be obtained even when the difference in SO
coupling for electron and hole excitations is neglected.
}
\label{fig0}
\end{figure}

In equilibrium, electrons occupy all quantum states inside the Fermi surface,
which for 2DEG is a circle in momentum space, centered at $\bm{p}=0$
and shown by a solid line in Fig.~\ref{fig0}a. However, if an electric
field is applied and finite current $\bm{j}$ flows in the system,
the electron distribution is shifted in momentum space by vector
$\Delta \bm{p}\simeq \pF(\bm{j}/j_{\rm F})$, where $j_{\rm F}=e\vF N_{\rm
e}$ and $N_{\rm e}=\pF^2/2\pi$ is the sheet density of 2DEG. In
this case electrons occupy all states within the dashed circle in
Fig.~\ref{fig0}a, centered at $\bm{p}=\Delta\bm{p}$.
The depleted states are called hole
excitations (holes) and the newly occupied states are
called electron excitations (electrons).

The net spin polarization
$\bar M$ of 2DEG is determined by the sum of
the electron and hole polarizations. Since these two polarizations
are directed in opposite directions, they mostly compensate each
other. However, for the linear in momentum SO coupling
$\hat H=\bm{p}\times \hat{\bm{\sigma}}/m^*\lambda_{\rm so}$,
the strength of the SO coupling is stronger for
electron excitations than for hole excitations, as illustrated in Fig.~\ref{fig0}a.
As the result, the magnitude of the spin polarization due to the SO coupling
of electrons a little bit exceeds that of holes. The net
polarization can be estimated as the difference in the energy
$\Delta E\simeq |\Delta\bm{p}|/m^*\lambda_{\rm so}$ of spin states
of electron  and hole excitations, multiplied by the
DoS $\nu_0=m^*/2\pi$. We find
\be
\frac{\bar M}{N_{\rm e}}\simeq
\frac{|\Delta \bm{p}|}{m^*\lambda_{\rm so}}
\frac{\nu_0}{N_{\rm e}}
\simeq
\frac{E_{\rm so}}{E_{\rm F}}\frac{j}{j_{\rm F}},
\label{estWF}
\ee
where $E_{\rm so}=\vF/\lambda_{\rm so}$.
From  Fig.~\ref{fig0}a we can also conclude that the vector of the spin
polarization is perpendicular to $\bm{j}$.

We note that the polarization is determined by the actual
current density $\bm{j}=\sigma_{\rm D}\check R(\wc\tautr)\bm{E}$
linear in the electric field $\bm{E}$, where
\be
\sigma_{\rm D} = \frac{e^2 \nu_0 \vF^2\tautr}{2},\quad
\check{R} (x)=\frac{1}{x^2+1}
\left[%
\begin{array}{cc}
  1 & -x \\
  x & 1 \\
\end{array}%
\right]
\label{sigmaD}
\ee
is the Drude conductivity tensor in the magnetic field $B\propto \wc$ and
$\tautr$ is the transport scattering time.
Consequently, if the current density $\bm{j}$ is fixed, the polarization
is independent of magnetic field. However, if the electric field $\bm{E}$
is fixed in the sample, then, according to \reqs{estWF} and \rref{sigmaD},
the polarization decreases and changes its orientation
as the magnetic field increases.

When the magnetic field becomes strong enough and $\wc\tauq\gtrsim 1$,
the electron DoS oscillates as a function
of energy, where $\tauq$ is the quantum scattering time.
As we discussed above, if a finite current
flows in 2DEG, the electron distribution is shifted in momentum
space. Now, due to the oscillations of the DoS,
see Fig.~\ref{fig0}b, the number of electrons and the number of
holes may be
different. In this case we can neglect the dependence of
SO coupling on the magnitude of the excitations' momentum and estimate
the spin polarization $\tilde M$ as the difference in the DoS of electron and
hole excitations, multiplied by the energy of SO splitting $E_{\rm
so}$, we have $\tilde M \simeq \delta \nu E_{\rm so}$. If electron
density of states oscillates with period $\wc$, we write
$\Delta \nu\propto \nu_0 \vF|\Delta p|/\wc$ and obtain
\be
\frac{\tilde M}{N_{\rm e}}\propto  \frac{E_{\rm so}}{\wc}\frac{j}{j_{\rm F}}.
\label{estSF}
\ee
Comparing \reqs{estWF} and \rref{estSF}, we conclude  that the spin
polarization due to the oscillations in the DoS contains the large
factor $E_{\rm F}/\wc\gg 1$, and therefore may be significantly
larger than the polarization at zero magnetic field.

In the above discussion we assumed that electron temperature is zero.
Temperature smearing of electron distribution function
does not affect the result of~\req{estWF}, calculated for the
constant DoS, but the estimate \req{estSF} for oscillating DoS
would contain difference of the DoS of electron and hole
excitations, averaged over the thermally smeared part of the
distribution function. This difference is suppressed if
temperature is higher than the period $\wc$ of oscillations of the
DoS. Thus,  \req{estSF} represents the upper limit for
the spin polarization in strong magnetic fields and the actual
polarization may be smaller. In the rest of this paper we present
the results of detailed analytical calculations of the spin
polarization  in weak and strong magnetic fields.

\section{Kinetic equation}
We consider a 2DEG with linear in momentum
SO coupling, placed in a perpendicular
magnetic field, when the filling factor $\nu= E_{\rm F}/\wc\gg 1$.
In this case we can use the quantum kinetic theory~\cite{VA03}
developed within the self-consistent Born approximation~\cite{Ando}.
We assume that
the correlation length of disorder $\xi$
is much longer than the Fermi wavelength $\lambda_{\rm
F}=2\pi/\pF$, and therefore the ratio of the
transport scattering time $\tautr$ to the quantum scattering time
$\tauq$ is large, $\tautr/\tauq\sim (\xi \pF)^2\gg 1$.
The kinetic equation has the form:
\be
\partial_t\hat f(t,\varepsilon)+i[\hat H;\hat f(t,\vare)]=
{\rm St}\{\hat f(t,\vare)\},
\label{kineq0}
\ee
where $[\hat A;\hat B ]=\hat A\hat B- \hat B\hat A$ and
\be
\hat H=
\frac{\hat{\bm{P}}^2}{2m^*}+e\bm{E}\bm{r}-\frac{\Delta_g}{2}\hat \sigma_z,\
\Delta_g=g\muB B
\label{H_General}
\ee
is the Hamiltonian of 2DEG in a quantum well with
\be
\hat{\bm{P}}=\bm{p}+\hat{\bm{\Lambda}}; \quad
\hat{\Lambda}_x=\frac{\hat\sigma_y}{\lambda_x}, \
\hat{\Lambda}_y=-\frac{\hat\sigma_x}{\lambda_y}.
\label{Lambda_vector}
\ee
Here $\bm{p}$ is the momentum operator and $\hat{\bm{\Lambda}}$
characterizes the SO coupling and contains
both the Rashba (the only term if $\lambda_x=\lambda_y$)
and crystalline anisotropy terms.
The second term in \req{H_General} describes the effect of the in-plane
electric field $\bm{E}$, and the last term represents the Zeeman energy,
$g$ is the electron gyromagnetic factor, $\muB=e/(2m_{\rm e}c)$
is the Bohr magneton, $m_{\rm e}$ is the free-electron mass.

In smooth disorder, the collision integral in \req{kineq0}
can be represented in the form, see Ref.~\cite{VA03}:
\be
{\rm St}\{\hat f(\vare)\}=
\frac{\nabla_{\hat{\bm{P}}}^2\{ \hat \nu(\vare);\hat f(\vare))\}
-\{\nabla_{\hat{\bm{P}}}^2 \hat\nu(\vare);\hat
f(\vare)\}}{2\nu_0\tautr}.
\label{st}
\ee
Here
$\nabla_{\hat{\bm{P}}}= \hat{\bm{P}}\times \partial_{\hat{\bm{P}}}$,
$\{\hat A;\hat B\}=\hat A\hat B+\hat B\hat A$, and
$\nu_0=m^*/(2\pi)$ is DOS per spin in zero magnetic field.
We neglected the Zeeman term in \req{st}. This approximation is
justified in weak magnetic fields, when the Zeeman energy is
small, as well as in strong magnetic fields, when the spin
orientation is fixed by the Zeeman field and
$\nabla_{\hat{\bm{P}}}$ can be replaced by  $\nabla_{\bm{p}}$.
Note that the collision integral is determined by the
electron DoS $\hat\nu(\vare)$~\cite{VA03}.

In a perpendicular magnetic field $B$,
the momentum operators $p_\alpha$
($\alpha=x,y$) do not commute:
\be
[p_\alpha;p_\beta]=-\frac{i}{\lambda_H^2}\epsilon_{\alpha\beta};
\ \ \
\hate  =\left[%
\begin{array}{cc}
  0 & 1 \\
  -1 & 0 \\
\end{array}%
\right];
\ \ \
\lambda_H=\sqrt{\frac{c}{eB}}.
\label{p_comm}
\ee
We represent the momentum operator $\bm{p}$ in the form
\be
\bm{p}=\pF\nn_{ \varphi}+\delta\bm{p},\quad
\delta\bm{p}=\frac{1}{2\Rc}\left\{
\begin{array}{c}
    ne^{i \varphi} + e^{-i \varphi}   n\\
  -i  ne^{i \varphi} + i e^{-i \varphi}   n \\
\end{array}
\right\},
\label{pnew}
\ee
where $\nnf=(\cos\varphi,\sin\varphi)$ and
$\Rc=\vF/\wc$ is the cyclotron radius.
To satisfy \req{p_comm}, the operators
$n$ and $ \varphi$ have to obey the following commutation
relation
\be
[ n; e^{i\varphi}]=e^{i\varphi}; \quad
 n\to -i\partial_\varphi .
\label{nphi_comm}
\ee
The integer eigenvalues of the operator
$n$ have the meaning of the Landau level indices.

In the representation \rref{pnew} of momentum operators $\bm{p}$,
we have
$\hat{\bm{P}}^2/2m^*=\wc n+\vF\nnf\hat{\bm{\Lambda}}+\hat H_{\rm
sc}$, where $\hat H_{\rm sc} =\delta\bm{p}\hat{\bm{\Lambda}}/m^*$
describes the asymmetry of the SO coupling between electron and hole
excitations, discussed in the previous Section and illustrated in
Fig.~\ref{fig0}a.
Below we show that only the term $\hat H_{\rm sc}$ couples the
spin and charge components
of the electron distribution function in weak (non-quantizing) magnetic fields.
We further simplify the
kinetic equation by performing an auxiliary transformation
$\hat {\cal U}_\delta=
\exp\{i\lambda_H^2(\delta\p\hate\hat{\bm{\Lambda}})\}$
of $\hat{\bm{P}}^2/2m^*$,
and keeping terms up to the second order in
$\delta\p\hate\hat{\bm{\Lambda}}$.
This transformation is an analogue of the unitary transformation
of the Hamiltonian
in zero magnetic field~\cite{fn1} and corresponds to a tiny rotation of
the momentum and spin states on ``angle''
$\lambda_H^2(\delta\p\hate\hat{\bm{\Lambda}}) \sim \lambda_{\rm F}/\lambda_{x,y}\ll
1$. Therefore, we neglect the transformation under $\hat {\cal U}_\delta$
of the electron distribution function, $\hat f$;
the spin operator, $\hat{\bm{\sigma}}$; and the Zeeman energy term.
This transformation is used
only to simplify  $\nabla_{\hat{\bm{P}}}$ in the collision integral
and $\hat H_{\rm sc}$, the latter in the new basis is
\be
\hat H_{\rm sc} =
\frac{\hat\sigma_z }{m^*\lambda_x\lambda_y}
\left(-2i\partial_\varphi+1\right).
\label{LorentsSO}
\ee
The spin-charge coupling, $\hat H_{\rm sc}$,
originates from the electron-hole asymmetry of the Hamiltonian
\req{H_General} (due to the difference in  velocities
of electrons and holes at distance $\delta\bm{p}$ from the Fermi
surface).
The factor $1/(m^*\lambda_x\lambda_y)$ can be
associated with the curvature of electron energy
bands in momentum space (cf. Refs.~\cite{Bc,Bc2DEG},
where the effect of the Berry curvature on motion in coordinate
space is considered). The above derivation of \req{LorentsSO}
was based on the representation \req{pnew},
defined at $B\neq 0$; the same
form  $\hat H_{\rm sc}$
is valid at $B=0$~\cite{ALG04}.

For a spatially homogeneous and stationary in time system, to
the lowest order in $\lambda_{\rm F}/\lambda_{x,y}$,
we obtain the following kinetic equation:
\be
\begin{split}
\partial_t \hat f+ & \wc\partial_\varphi\hat f  +
i\left[\vF \nnf\hat{\bm{\Lambda}}-  \frac{\Delta_g}{2} \hat \sigma_z;\hat f\right ]
\\ &+i[\hat H_{\rm sc};\hat f]
+e\vF \nnf \bm{E}\partial_\vare \hat f
=\frac{\{\hat \nu(\vare);\partial_\varphi^2\hat
f\}}{2\nu_0\tautr},
\end{split}
\label{kineq1}
\ee
where function $\hat f(\vare,\varphi)$  describes the
distribution of electrons with momentum $\p$
in the direction $\nnf$. The second term $i\wc [n; \hat f]=
\wc\partial_\varphi \hat f$
in the left hand side of
\req{kineq1} describes the Lorentz force acting on electrons in
magnetic field $B\propto\wc$. The fourth term
$i[\hat H_{\rm sc};\hat f]$ with $\hat H_{\rm sc}$ given by
\req{LorentsSO} has a similar structure and can be associated with
the Lorentz force, induced by the SO coupling.
Below we solve \req{kineq1} in the limits of weak
($\wc\tauq\lesssim 1$) and strong ($\wc\tauq\gtrsim 1$) magnetic fields.

\section{Weak magnetic field}
At $\wc\tauq\ll 1$, the oscillatory
component of the DOS $\hat \nu(\vare)$ is exponentially
suppressed and $\hat\nu(\vare)=\hat 1\nu_0$.
We solve the kinetic equation \req{kineq1} by consecutive
iterations, limiting our consideration to the limit of weak SO coupling,
$\lambda_{x,y}\gg \vF\tautr$. We start with the Fermi distribution function
$\hat f(\vare)=\hat 1f_{\rm F}(\vare)$,
$f_{\rm F}(\vare)=(\exp((\vare-E_{\rm F})/T)+1)^{-1}$. To first order in
the electric field $\bm{E}$
the distribution function contains an anisotropic
component with respect to the momentum direction $\varphi$:
\be
\hat f^{(1)}(\vare,\varphi)=-\hat 1
\frac{2\sigma_{\rm D}}{e\vF\nu_0}
\left(\nnf^T\check{R}(\wc\tautr)\bm{E} \right)
f'_{\rm F}(\vare),
\label{f1}
\ee
where $\sigma_{\rm D}\check{R}(\wc\tautr)$ is the Drude conductivity matrix
\req{sigmaD}.

The distribution function $\hat f^{(1)}(\vare,\varphi)$
has no spin components. The spin components in $\hat f$
appear only if the spin-charge coupling term $i[\hat H_{\rm sc};\hat f ]$
is taken into account in \req{kineq1}. Keeping
$i[\hat H_{\rm sc};\hat f^{(1)}(\vare,\varphi) ]$
with $\hat H_{\rm sc}$ and $\hat f^{(1)}(\vare,\varphi)$
given by \reqs{LorentsSO} and \rref{f1},
we obtain  the solution of \req{kineq1} after the second iteration
in the form:
\be
\hat f^{(2)}(\vare,\varphi)=-\hat\sigma_z
\frac{4\sigma_{\rm D}\tautr}{e\vF\nu_0}
\frac{[\nnf^T\hate \check R^2(\wc\tautr)\bm{E}]}
{m^*\lambda_x\lambda_y}
f'_{\rm F}(\vare).
\label{f2}
\ee
Still, $\hat f^{(2)}(\vare,\varphi)$ does not describe
spin polarization of 2DEG because
$\int d\varphi \hat f^{(2)}(\vare,\varphi)=0$.
Substituting  $\hat f^{(2)}(\vare,\varphi)$
into the second term in the left hand side (LHS) of \req{kineq1},
we look for a solution
$
\hat f^{(3)}(\vare,\varphi)=\bm{f}^{(3)}(\vare)\bm{\sigma}+
\bm{\alpha}(\vare)\nnf\hat\sigma_z
$.
Here $\bm{f}^{(3)}(\vare)\bm{\sigma}=\sum_{j=x,y}f_j\hat\sigma_j$
is an isotropic spin term, which determines the polarization of
2DEG. First, we express
$\bm{\alpha}(\vare)\nnf $ in terms of $\bm{f}^{(3)}(\vare)$,
then, we insert $\bm{\alpha}(\vare)\nnf $
into the second term in LHS of \req{kineq1} and average
the result over $\nnf$.
This procedure is equivalent to neglecting higher harmonics in
$\nnf$,  small in the parameter $\vF\tautr/\lambda_{x,y}\ll 1$.
We find
\be
\label{Kmatrix}
\begin{split}
&\bm{f}^{(3)}(\vare)  =\kappa
\check K(\wc\tautr)\bm{g},\ \ \
\kappa=\frac{\lambda_x\lambda_y}{2\vF^2\tautr},
\\
& \hat K(x)  = \frac{\displaystyle
\left[%
\begin{array}{cc}
   \lambda_x/\lambda_y & x(1+\eta(1+x^2) ) \\
  -x(1+\eta(1+x^2) ) &   \lambda_y/\lambda_x  \\
\end{array}%
\right]}
{(x^2(1+\eta(1+x^2))^2+1)(1+x^2)^{-1}}.
\end{split}
\ee
The vector $\bm{g}$ is the spin generation matrix
$\bm{g}\hat{\bm{\sigma}} =
i \vF \int\nnf [\hat{\bm{\Lambda}}; \hat f^{(2)}(\vare,\varphi)]
d\varphi/2\pi$,
the matrix $(\kappa \check K)^{-1}$ is the spin relaxation matrix,
and $\eta=g(m/m_{\rm e})(\lambda_x\lambda_y/4\vF^2\tautr^2)$
characterizes the strength of the Zeeman splitting.

The ratio of spin density
$\bm{M}=\nu_0\int \bm{f}^{(3)}(\vare)d\vare$
to the total electron density
$N_{\rm e}=\pF^2/2\pi$ of 2DEG is
\be
\begin{split}
\frac{\bm{M}}{N_{\rm e}} & =  {\cal M}_0\bm{m}
,
\quad
{\cal M}_0 =\frac{\lambda_{\rm F}^2}{(2\pi)^2}
\frac{e |\bm{E}| \tautr}{\sqrt{\lambda_x\lambda_y}},
\\
\bm{m} & =
\check K\check L\hate\check R^2(\wc\tautr)\bm{e},
\quad
\check L=\left[%
\begin{array}{cc}
   \sqrt{\frac{\lambda_y}{\lambda_x}} & 0 \\
  0 & \sqrt{\frac{\lambda_x}{\lambda_y}} \\
\end{array}%
\right],
\end{split}
\label{noSdH}
\ee
where $\lambda_{\rm F}=2\pi/\pF$ is the Fermi wavelength,
length scales $\lambda_{x,y}$ describe the strength of the SO
coupling, \req{Lambda_vector}, and matrices $\hat R$ and $\hat K$
are introduced in \reqs{sigmaD} and \rref{Kmatrix}.

\begin{figure}
\epsfxsize=0.40\textwidth
\centerline{\epsfbox{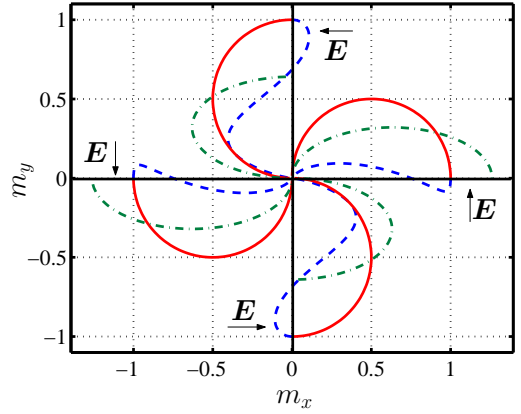}}
\caption{(Color online) A parametric dependence of the polarization vector
on the orbital magnetic field $B\propto \wc\tautr$  is shown for different
values of $\eta$ and $\lambda_x/\lambda_y$. The curves
represent the Rashba coupling ($\lambda_x=\lambda_y$)
with $\eta=0$ [solid line] and $\eta=1$ [dashed line]. The
dash-dotted line represents $\lambda_x=2\lambda_y$ and $g=0$.
}
\label{figurexy}
\end{figure}

The direction of the spin polarization $\bm{M}$ is given
by the vector $\bm{m}$, which is related to the direction of the
electric field $\bm{e}=\bm{E}/|\bm{E}|$ through
the tensor $\check K\check L\hate\check R^2$.
For the Rashba coupling,
$\lambda_{\rm so}=\lambda_{x,y}$,  \req{noSdH}
coincides with the result of Ref.~\cite{ALG1,MSH04} at $B=0$, obtained for point-like
scatters, if the full scattering time is replaced by $\tautr$.
We can reduce \req{noSdH}
in case $\eta\propto g=0$ and $\lambda_{\rm so}=\lambda_{x,y}$ to
$\bm{M}/N_{\rm e}=(E_{\rm so}/E_{\rm F})(\hate\bm{j}/j_{\rm F})
$, where
$\bm{j}=\sigma_{\rm D}\check R(\wc\tautr)\bm{E}$ is the electric
current density, $E_{\rm so}=\vF/\lambda_{\rm so}$
is the SO energy splitting,
and $j_{\rm F}=e\vF N_{\rm e}$ (the current density
if all electrons were moving with velocity $\vF$). In typical
2DEG, $E_{\rm so}\ll E_{\rm F}$ and $|\bm{j}|\ll j_{\rm F}$, thus
the polarization is small:
$|\bm{M}|/N_{\rm e}\ll 1$. Note that for fixed $\bm{j}$,
$\bm{m}$ is independent of $B$.
Finite $g$ factor and anisotropy of SO coupling
($\lambda_x\neq\lambda_y$) do not significantly change  the
value of $|\bm{M}|/N_{\rm e}$, but result in more complicated
behavior of $\bm{m}$ as a function of $B$.
For fixed $\bm{E}$, we show dependence of $|\bm{m}|$  on
$B\propto\wc$ in Fig.~\ref{fig1} and the parametric plot
of $\bm{m}(B)$ in $(m_x,m_y)$ plane  in Fig.~\ref{figurexy}.

\section{Strong Magnetic field}
As the magnetic field increases, the polarization \req{noSdH}
decreases. However,  
if $\wc\tauq\gtrsim 1$,
DOS becomes an oscillating function of energy shifted
in opposite directions for different spin states. The magnitude of the shift is
equal to either the Zeeman energy $\Delta_g$, \req{H_General},
or the SO energy $\Delta_{\lambda}$, see \req{Delta} below.

We first discuss the case of strong magnetic fields
$B\gtrsim B_*$,
when $\Delta_g\gg\Delta_{\lambda}$,
$B_*=(c/e)\sqrt{m_{\rm e}/gm^*}/\lambda_{x,y}\lambda_{\rm F}$ and
the splitting of the spin states is dominated by the Zeeman
effect.
In strong magnetic fields we can neglect the effect of the SO
coupling on the electron DoS. In this case we can obtain the DoS
for the two spin projections $\sigma_z/2$ on magnetic field
independently from each other following the derivation in Ref.~\cite{VA03}
and taking into account the Zeeman splitting $\Delta_g=g\muB B$:
\be
\nu_{\sigma_z}=
\nu_0\left[
1+2\sum_{l=1}^{\infty} (-\delta)^l g_l \cos\frac{2\pi l
(\vare-\sigma_z\Delta_g/2)}{\wc}
\right],
\label{DoS}
\ee
where the spectral coefficients $g_l=L_{l-1}^1(2\pi l/\wc\tau_{\rm q})/l$
are expressed in terms of the Laguerre polynomials $L_n^m(x)$, and
$\delta =\exp(-\pi/\wc\tau_{\rm q})$.

The electric field $\bm{E}$ generates an anisotropic component of
the electron distribution function:
$\hat f^{(1)}(\vare,\varphi)= \hat 1f^{(1)}_{\rm c}(\vare,\varphi)
+\hat f^{(1)}_{\rm s}(\vare,\varphi)$.
Calculations of $\hat f^{(1)}(\vare,\varphi)$ are similar to those
that lead to \req{f1}.
Now, in addition to the charge contribution
$f^{(1)}_{\rm c}(\vare,\varphi)$,
the distribution function contains the spin component
$\hat f^{(1)}_{\rm s}(\vare,\varphi)$:
\be
\hat f_{\rm s}^{(1)}=\frac{ 2e\vF\hat\sigma_z} {\wc^2\tautr}
\nnf\bm{E} f'_{\rm F}(\vare)
\sum\limits_{l=1}^{\infty}(-\delta)^l g_l K_l
\sin\frac{2\pi l\vare}{\wc},
\label{f1z}
\ee
where $K_l=\sin (2\pi l \Delta_g/\wc )$.
Due to the oscillations of the DOS we already generated
a spin component $\hat f^{(1)}_{\rm s}(\vare,\varphi)$
after the first iteration of \req{kineq1}. In weak magnetic
fields, this spin component is exponentially small, and
to obtain spin polarization, we have to
take into account finite curvature of the electron spectrum on the
scale of energy band $E_{\rm F}$, described by
the term $\hat H_{\rm sc}$, see
\req{LorentsSO}.
If $\wc\gtrsim 1/\tauq$, the particle-hole asymmetry appears
on energy scale $\wc$ and we find the spin components in
$\hat f$ without taking into account $\hat H_{\rm sc}$:
the component $\hat f_{\rm s}^{(1)}(\vare,\varphi)$ is already
similar in its properties to $\hat f^{(2)}(\vare,\varphi)$.
To calculate the polarization we just follow the procedure
described below \req{f2} using
$\hat f^{(1)}_{\rm s}(\vare,\varphi)$ instead of
$\hat f^{(2)}(\vare,\varphi)$. Since
$\wc\tauq\gtrsim 1$, we take $\wc\tautr=x\gg 1$
in \req{Kmatrix} for $\check K(x)$. Substituting
$\hat f^{(3)}(\vare,\varphi)$
to the expression for the polarization
$
\tilde{\bm{M}}=\int  {\rm Tr}(\hat{\bm{\sigma}}
\{\hat\nu(\vare); \hat f^{(3)}\})d\vare d\varphi/(8\pi)
$,
we obtain
\be
\begin{split}
\frac{\tilde{\bm{M}}}{N_{\rm e}} & =
{\cal M}_g\frac{\check L\bm{E}}{|\bm{E}|}
\sum\limits_{l=1}^{\infty} (-1)^l\zeta_l {\cal Y}\left(\frac{lT}{\wc}\right)
\sin 2\pi l \nu,
\\
{\cal M}_g & =
\frac{2\lambda_H^2}{(\wc\tautr)^2}
\frac{e|\bm{E}|\tautr}{\sqrt{\lambda_{x}\lambda_{y}}}, \quad
{\cal Y}(x)=\frac{2\pi^2x}{\sinh2\pi^2x},
\label{Mosc}
\end{split}
\ee
the amplitudes $\zeta _l$ are given by
\be
\zeta_l\!\!=\!\! \sum_{k=-\infty}^{+\infty}\!\!
\delta^{|k|+|k+l|}g_{|k|}g_{|k+l|}
\frac{\wc}{\Delta_g} \sin\frac{\pi k\Delta_g}{\wc}
\cos\frac{\pi |k+l|\Delta_g}{\wc},
\label{zetal}
\nonumber
\ee
$\nu=E_{\rm F}/\wc$ and $\Delta_g/\wc=gm^*/2m_{\rm e}$.
We notice that  $\tilde{\bm{M}}$
oscillates as a function of $\nu$ and is exponentially
suppressed if $\wc\lesssim T$ or
$\wc\lesssim 1/\tauq$. Thus, the conditions for observation
of the oscillating component
of the polarization are similar to those for observation of
the Shubnikov--De Haas oscillations in
the conductivity, cf. \req{Mosc} to Eq.~(4.18) in Ref.~\cite{VA03}.

Next, we consider the range of magnetic fields,
$B\lesssim B_*$,
when the spin component in the DOS is created due to
the SO splitting of electron states with opposite helicity.
Taking into account the SO coupling in
the original basis, where $\hat \nu(\vare)=\hat \nu(\vare,\varphi)$
is non-diagonal in spin space and depends on the momentum direction
$\nnf$ is cumbersome.
The calculations become easier in the rotated basis defined for
$\Rc/\lambda_{x,y}\ll 1$ by the matrix
$
\hat{\cal U}_0=\exp\{i\Rc(\nnf\hat\epsilon
\hat{\bm{\Lambda}})\},
$
(for $\lambda_{\rm R}=\lambda_{x,y}$ this rotation
can be used for arbitrary $\Rc/\lambda_{\rm R}$).
In the rotated basis, the spectral function $\hat \nu(\vare)$
is isotropic and is given by \req{DoS} with $\Delta_g$ replaced by
$\Delta_\lambda$:
\be
\Delta_\lambda=\frac{2\vF^2}{\lambda_x\lambda_y}\frac{1}{\wc},
\quad \frac{\vF}{\wc}\ll\lambda_{x,y}.
\label{Delta}
\ee
The kinetic equation \req{kineq1} for the rotated electron
distribution function
$\hat {\cal F}=\hat{\cal U}_0^\dagger \hat f \hat{\cal U}_0$
is also modified:
\be
\begin{split}
[\partial_t +
\wc\partial_\varphi ]\hat {\cal F} &  -
\frac{i\Delta_\lambda}{2} [\hat\sigma_z,\hat {\cal F}]
+e\vF \nnf \bm{E}\partial_\vare \hat {\cal F}
={\widetilde {\rm St}}[\hat {\cal F}],
\end{split}
\label{kineq2}
\ee
where
the collision integral has the form
\begin{subequations}
\be
{\widetilde {\rm St}}[\hat {\cal F}]  =
\frac{\tilde \partial_\varphi^2 \{\hat \nu(\vare);\hat {\cal F}\} -
\{ \tilde \partial_\varphi^2 \hat \nu(\vare);\hat {\cal F}\}}
{2\nu_0\tautr},
\label{Strotated}
\ee
with
\be
\tilde \partial_\varphi \hat {\cal F}=\partial_\varphi \hat {\cal F}-
i\left[\Rc\nnf\hat{\bm{\Lambda}} +
\frac{\hat\sigma_z \Delta_\lambda}{2\wc};\hat {\cal F}\right].
\ee
\end{subequations}

For simplicity, we consider the limit $\wc\tauq\lesssim
1$, and keep terms linear in $\delta$
[if there is a window in $\wc$ where  $\wc\tauq\gtrsim 1$ and
$\Delta_\lambda\gg \Delta_g$, exact DOS has to be used,
cf. \req{Mosc}].
Then, the contribution
to the distribution function due to electric field $\bm{E}$
is given by
$\hat {\cal F}^{(1)}(\vare,\varphi)=
\hat 1 {\cal F}^{(1)}_{\rm c}(\vare,\varphi)
+\hat {\cal F}^{(1)}_{\rm s}(\vare,\varphi)$.
$\hat {\cal F}^{(1)}_{\rm s}(\vare,\varphi)$ has the form of
\req{f1z}, with only one term $l=1$,
and $K_1^g$ replaced by
$\sin(2 \pi \Delta_\lambda/\wc)$.
Substituting  the spin component
$\hat {\cal F}^{(1)}_{\rm s}(\vare,\varphi)$ to the collision integral
in  \req{kineq2} with $\hat\nu(\vare)=\hat 1\nu_0$
(oscillating components in $\hat\nu(\vare)$
produce extra factor $\delta\ll 1$), we obtain an isotropic in $\nnf$
spin component of the electron distribution function.
To the lowest order in $\Rc/\lambda_{x,y}$ and $(\wc\tautr)^{-2}$,
the polarization is
\be
\begin{split}
\frac{\tilde{\bm{M}}}{N_{\rm e}} & =
 {\cal M}_\lambda \left[-\hate \check L\hate \bm{e}\right]
e^{-\pi/\wc\tauq}{\cal Y}\left(\frac{T}{\wc}\right)\sin 2\pi\nu , \\
{\cal M}_\lambda   & = \frac{4\pi\lambda_H^2}{(\wc\tautr)^3}
\frac{ e|\bm{E}|\tautr}{\sqrt{\lambda_{x}\lambda_{y}}}.
\label{Mso}
\end{split}
\ee
Both Zeeman  and SO splitting of DOS
result in qualitatively similar expressions for the oscillatory
polarization, cf. \req{Mosc} and \rref{Mso}; depending on the relation between
$\Delta_g$ and $\Delta_\lambda$, either \req{Mosc} or \rref{Mso} is
applicable.

\section{Discussions and Conclusions}

First, we notice that the kinetic equation
approach developed here can be further generalized
to describe non-stationary in
time systems. For illustration, we consider
the homogeneous spin relaxation in 2DEG placed in a perpendicular magnetic field,
recently studied both theoretically~\cite{BB04} and experimentally~\cite{Sih}.
For the in-plane spin polarization
$\hat f_\| =f_x\hat\sigma_x+f_y\hat\sigma_y$ in
non-quantizing magnetic field $\wc\tauq\ll 1$ we have
\be
\frac{\partial \hat f_\|}{\partial t}=-
\frac{2\vF^2\tautr}{\lambda_x\lambda_y}
\hat K^{-1}(\wc\tautr)\hat f_\|,
\label{fperp}
\ee
where $\hat K^{-1}$ is the inverse matrix of $\hat K$
defined by \req{Kmatrix}. The off-diagonal elements of $\hat K^{-1}$
describe spin precession due to the Zeeman field and the SO
coupling. In strong magnetic fields $B\gg B_*$, only the Zeeman
component survives, however, in weaker fields, $B\ll B_*$, the
dominant contribution to the precession rate originates from SO
coupling.
For the polarization $\hat f_\bot=f_z\hat\sigma_z$
perpendicular to 2DEG we find
\be
\frac{\partial f_z}{\partial t}=-
\frac{2\vF^2\tautr}{1+\wc^2\tautr^2}
\left[\frac{1}{\lambda_x^2}+\frac{1}{\lambda_y^2}
\right]f_z
\label{fz}
\ee

The structure of \reqs{fperp} and \rref{fz} is
consistent with the
result of Ref.~\cite{BB04}, obtained for short range
disorder,
when the quantum scattering time $\tau_{\rm q}$ and
the transport scattering $\tau_{\rm tr}$ are equal. For long range
disorder the spin relaxation is governed by the transport
scattering time. Thus, scattering processes
with large change of electron momentum are responsible for spin relaxation.

In this paper we demonstrated that in sufficiently strong
magnetic fields, $\wc\gg \{T,\ 1/\tauq\}$, the
factors $\zeta_l$ and ${\cal Y}(x)$ in \req{Mosc}
become of  order of unity. Then, the ratio of
the amplitude of the oscillatory polarization $\tilde{\bm{M}}$,
\req{Mosc}, to the polarization $\bm{M}$ at $B=0$,
\req{noSdH}, is characterized by
\be
\frac{{\cal M}_g}{{\cal M}_0}\propto \frac{E_{\rm
F}}{\wc}\frac{1}{(\wc\tautr)^2}, \quad {\rm for}\ \ \ \wc\tauq\gtrsim 1
\label{MgM0}
\ee
The large
factor $E_{\rm F}/\wc\gg 1$ is related to the enhancement
of the electron-hole asymmetry by magnetic field on energy scale $\wc$.
The factor $(\wc\tautr)^{-2}$ describes the suppression of
the diffusion coefficient by magnetic field.

One can expect that the magnitude of the polarization,
which is linear in the applied electric field, can be increased
significantly by applying stronger electric field.
However, in experiments
the magnitude of the electric field $\bm{E}$ is limited
by the heat dissipation
$\sigma_{\rm D}\bm{E}^{\rm T}\hat R(\wc\tautr)\bm{E}\propto
(\wc\tautr)^{-2}$ in the sample. Because this heat power is
also suppressed at $\wc\tautr\gg 1$, one can apply stronger
electric field $\bm{E}$ to partially compensate
the factor $(\wc\tautr)^{-2}$ in \req{MgM0}.
Thus, the amplitude of the oscillatory polarization, achievable in
experiments, could exceed the polarization in zero magnetic field
even by larger factor $E_{\rm F}\tauq^2/\tautr$ than
the estimate \req{MgM0}.

We discussed the behavior of the current--induced
spin polarization of 2DEG. This phenomenon is only one of the examples
of the magneto-electric effect, originating in materials with
spin-orbit coupling.
Other magneto-electric effects, such as the photocurrent
induced by optical orientation
of electrons~\cite{ILGP}, can be enhanced by a quantizing magnetic
field as well.

I would like to thank I. Aleiner, Y. Alhassid, S. Girvin,
Yu. Lyanda-Geller and E. Mishchenko
for stimulating discussions and comments. This work was supported by the
W. M. Keck Foundation and by NSF Materials Theory grant DMR-0408638.


\end{document}